\documentclass[aps,pre,amsmath,amssymb,preprint,
superscriptaddress,
]{revtex4-1}
\ifx\pdfoutput\undefined
\usepackage{graphicx}
\else
\usepackage{graphicx}
\usepackage{epstopdf}
\fi
\usepackage[pdftitle={Yongxiang HUANG},bookmarks=true,citecolor=red,colorlinks=false]{hyperref}
\usepackage{wasysym}
\usepackage{xcolor}
\usepackage{siunitx}

\usepackage{CJK}
\newcommand{\upd}{\mathrm{\,d}}

\newcommand{\red}[1]{\textcolor{black}{#1}}

\begin{document}

\begin{CJK*}{GB}{gbsn} 
\title{Intrinsic flow structure and multifractality in two-dimensional bacterial turbulence}

\author{Lipo Wang (ÍõÀûÆÂ)}%
\email{lipo.wang@sjtu.edu.cn}
\affiliation{UM-SJTU Joint Institute, Shanghai JiaoTong University, Shanghai 200240, RP China}

\author{Yongxiang Huang (»ÆÓÀÏé)}
\email{yongxianghuang@gmail.com}
\affiliation{State Key Laboratory of Marine Environmental Science, College of Ocean and Earth Sciences, Xiamen University, Xiamen 361102, PR China}

\date{\today}

\begin{abstract}
The active interaction between the bacteria and fluid generates turbulent structures even at zero Reynolds number. Velocity of such a flow obtained  experimentally has been quantitatively investigated based on streamline segment analysis. There is a clear transition at about $16$ times of the organism body length separating two different scale regimes, which may be attributed to the different influence of the viscous effect. Surprisingly the scaling extracted from the streamline segment indicates the existence of scale similarity even at the zero Reynolds number limit. Moreover the multifractal feature can be quantitatively described via a lognormal formula with the Hurst number $H=0.76$ and the intermittency parameter $\mu=0.20$, which is coincidentally in agreement with the three-dimensional hydrodynamic turbulence result. 
The direction of cascade is measured via the filter-space technique. An inverse energy cascade is confirmed. For the enstrophy, a forward cascade is observed when $r/R\le 3$, and an inverse one is observed when $r/R>3$. Additionally, the lognormal statistics is verified for the coarse-grained energy dissipation and enstrophy, which supports the lognormal formula to fit the measured scaling exponent.
\end{abstract}

\maketitle
\end{CJK*}
\section{Introduction}
In three-dimensional (3D) hydrodynamic turbulence scale invariant properties are inherited at different scales via the cascade process, i.e. the Richardson-Kolmogorov energy cascade,  where energy is transferred from larger to smaller scales until dissipated to heat at viscosity scale~\citep{Frisch1995}. It is generally believed that the energy injection is through the mean flow or the large-scale movement.
To characterize such complex dynamics at different scales, the structure function based on velocity increment has been widely applied in different turbulent systems, including bacterial turbulence \citep{Wensink2012PNAS}. Conventionally the structure function is defined as
\begin{equation}
 S_q(r)=\langle  \Delta  \mathbf{u}_r(\mathbf{x})^q\rangle_{\mathbf{x}}\sim r^{\zeta(q)},\,\Delta \mathbf{u}_r(\mathbf{x})=\mathbf{u}(\mathbf{x}+\mathbf{r})-\mathbf{u}(\mathbf{x}).
\label{velocity}
\end{equation}
Here $\mathbf{x}$ represents the spatial coordinate and $r=\vert \mathbf{r}\vert$ is the separation scale lying in the inertial range $\ell_{\nu}\ll r\ll L$, where $\ell_{\nu}$ is the Kolmogorov scale and $L$ is the integral scale.  The existing results \citep{Davidson2005PRL,Huang2011PRE,Schmitt2016Book} show that the aforementioned definition in equation (\ref{velocity}) could mix information from different flow structures; thus it is difficult to detect the scaling index $\zeta(q)$.

In general, the scale $r$ is determined by the dynamic process itself, for instance the topology of the fluid structure. In recent years several new approaches have been proposed to overcome this difficulty, e.g., detrended fluctuation analysis \citep{Peng1994PRE},  Hilbert-Huang transform \citep{Huang1998EMD,Huang2008EPL} and multi-level segment analysis \citep{Wang2015JSTAT}. For the hydrodynamic turbulent system, the measured $\zeta(q)$ is nonlinear with respect to $q$, which is termed as multifractality. Physically, multifractality originates
from the nonlinearity of the Navier-Stokes equation.

In active flows the living matter such as bacteria interact with the fluid. Thus the patterns of energy injection and dissipation are different from the classic Navier-Stokes governed flows, raising more complexities to be investigated \citep{Thampi2013PRL,Thampi2014PTRS,Fodor2016PRL}. For example, in bacterial turbulence the energy injection scale is comparable to $R$, the body length scale of the microorganisms. $R$ is often few or dozens of $\mu$m, which is much smaller than the fluid dissipative scale  $\ell_{\nu}$. This interesting flow status has been redefined as ``mesoscale turbulence'' in living fluid \citep{Wensink2012PNAS,Bratanov2015PNAS}. Although the background flow has much smaller Reynolds numbers than that required for conventional fluid turbulence, under the self-propulsion action by the microorganisms, the flow still exhibits a turbulent-like movement even at zero Reynolds number
\citep{Wu2000PRL,Pooley2007PRL,Ishikawa2008PRL,Rushkin2010PRL,Ishikawa2011PRL,
Chen2012PRL,Wensink2012PNAS,Dunkel2013PRL,Saintillan2012JRSI,Dunkel2013NJP,Grobmann2014PRL,Qiu2016PRE}. Meanwhile the effects from the flow can be important not only  for nutrient mixing, information passage and thus the biological behavior, but also for driving  micromachines \citep{Thampi2016science}.

Experimentally the  velocity field in bacterial turbulence can be measured in the Eulerian frame via the confocal particle image velocimetry (PIV) technique \citep{Ishikawa2011PRL,Chen2012PRL}. Flow visualization shows clearly that the collective motion of bacteria in the suspension exhibits coherent structures on a scale much larger than the individual body length, e.g. 10 times $R$ \citep{Wensink2012PNAS}. This important property can be understood from a simplified picture of interaction between  individual swimmers \citep{Pooley2007PRL,Ishikawa2008PRL,Grobmann2014PRL,Saintillan2012JRSI}. Theoretically there are different models to predict the active fluids by constructing the possible field governing equations ,inevitably with some control parameters \citep{Dunkel2013PRL,Wensink2012PNAS,Dunkel2013NJP,Fodor2016PRL}. Numerical tests show that by adjusting these control parameters some of the measurement results can satisfactorily be reproduced, for instance the large scale coherent structure \citep{Dunkel2013PRL,Bratanov2015PNAS}. Wensink et al., \citep{Wensink2012PNAS} observed a dual-power-law behavior from the solution of a two-dimensional (2D) bacterial turbulence model equation. Recently, Qiu et al., \citep{Qiu2016PRE} confirmed the intermittency correction in bacterial turbulence via a Hilbert-based methodology. From the observed dual-power-law behavior, it can be estimated that intermittency in the smaller scale regime is stronger than that in the large scale regime. In spite of these latest progresses, the complexities of flow and bacteria interaction are still scarcely understood.

\section{Data Presentation}
\begin{figure}[!htp]
\centering
\includegraphics[width=0.85\linewidth,clip]{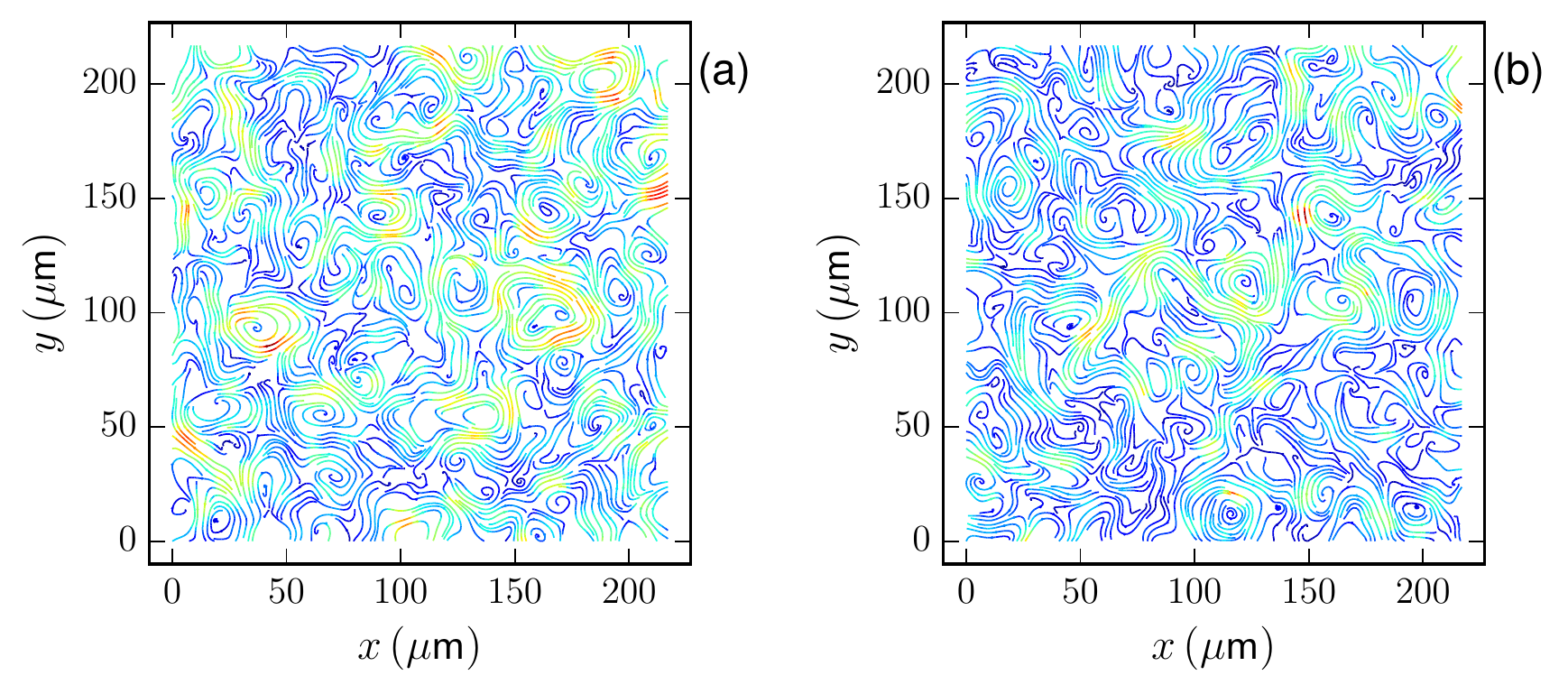}
  \caption{(Color online) Two snapshots of the instantaneous streamline colored with the velocity magnitude. Visually the flow field is smooth with a typical  large-scale structure size of $\sim 50\,\si{\mu m}$.}\label{fig:snapshot}
\end{figure}

\begin{figure}[!htp]
\centering
\includegraphics[width=0.65\linewidth,clip]{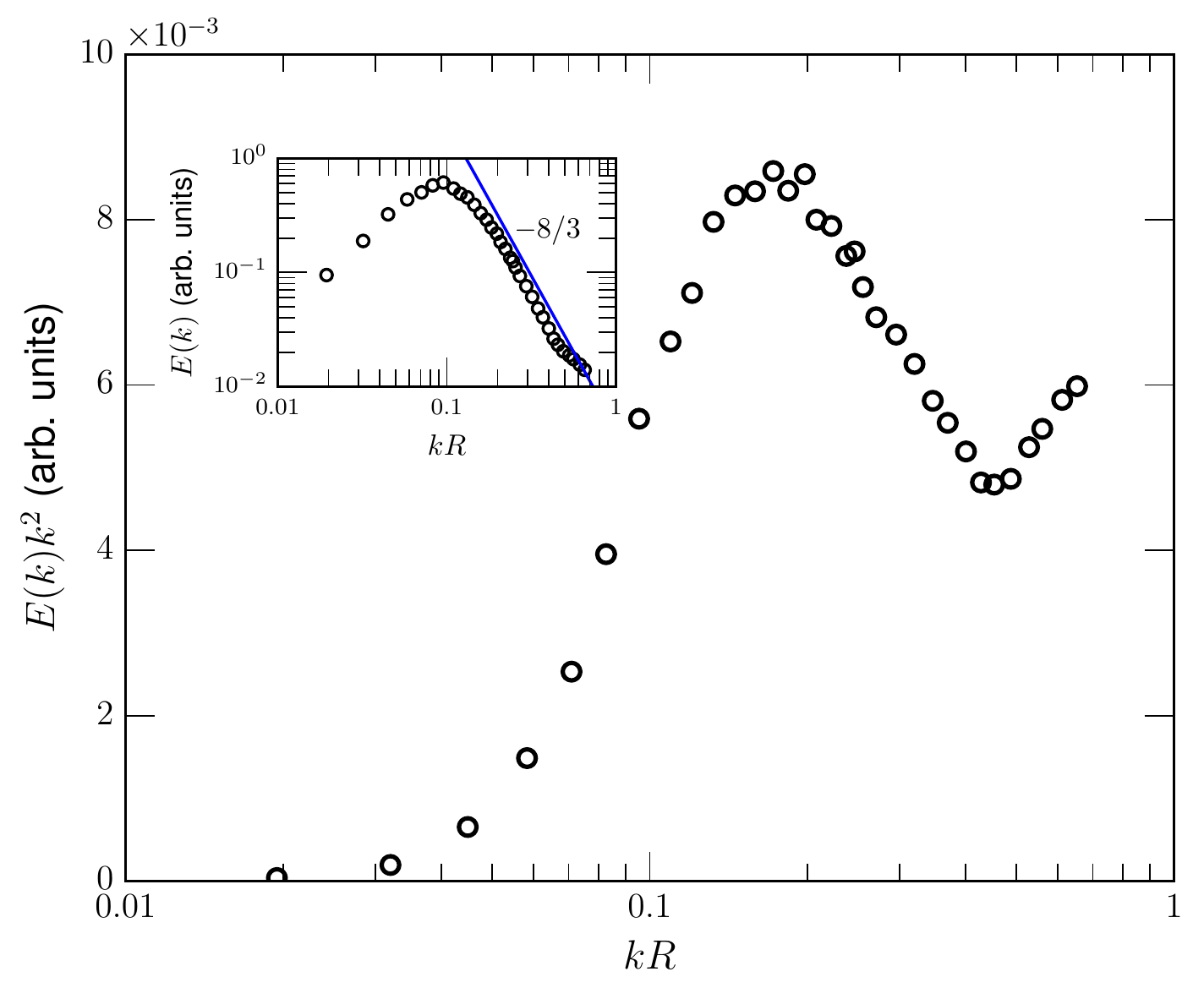}
  \caption{(Color online) \red{Experimental dissipation spectrum $E(k)k^2$, where a peak is found to be around $k_pR=0.17$, corresponding to a spatial scale $\ell_p\simeq 6R$. The inset shows the Fourier power spectrum $E(k)$, where a power-law fitting $-8/3$ on the range $ 0.2\le kR\le 0.4$ is illustrated as a solid line. }}\label{fig:dissipation}
\end{figure}

The data analyzed here is from the experimental results by courtesy of R.~E. Goldstein. We recall briefly the
main control parameters in a microfluidic chamber, which has a vertical
height $H_c$ less or equal to the individual body length $R$.
For bacterial suspension in a thin fluid, the spatial scale of the flow
structure is much larger than the depth and thus the system can be
approximated as 2D. The Bacillus subtilis has an individual
body length $R\simeq 5\,\mu$m and an aspect ratio $a=5$, i.e. the ratio
between the body length $R$ and the body diameter $d$. The kinetic
energy is injected into the system by the living matter at
approximately the scale $R$. The volume filling fraction is $\phi=84\%$
with a total particle number $N=9968$, to ensure the turbulent phase of the flow \citep{Wensink2012PNAS}. The PIV measurement area is $217\mu\mathrm{m}\times 217\mu\mathrm{m}$, and the image resolution is of $700\,$pix$\times700\,$pix with conversion rate $0.31\,\mu$m/pix and frame rate  $40$Hz. The commercial PIV software Dantec Flow Manager is used to extract the flow field component with a moving window size $32\,$pix$\times 32\,$pix and $75\%$ overlap, which results a  $84\times 84$ velocity vectors and a total 1441 snapshots, corresponding to a time period $\sim36\,$seconds. Totally, there are 10,167,696 velocity vectors.  \red{The mean and root-mean-square (rms.) velocities were determined as $
\overline{\mathbf{u}}=(0.36,0)\,\mathrm{\mu ms^{-1}}$ and $\mathbf{u}^{'}=(0.30,0.29)\,\mathrm{\mu ms^{-1}}$, respectively, and the
corresponding turbulent intensity was around $\mathbf{u}'/\overline{\mathbf{u}}_x\simeq 82\%$. In the following analysis, the mean velocity is removed from the velocity field.}

Figure \ref{fig:snapshot} shows two snapshots of the instantaneous streamlines colored with the velocity magnitude. At different instants the typical field structure can be clearly observed with a spatial size around $\sim 50\,\si{\mu m}$, corresponding to $10R$. \red{Figure \ref{fig:dissipation} shows the so-called dissipation spectrum $E(k)k^2$ \citep{Monin1971,Ishikawa2011PRL}, where $E(k)$ is the Fourier power spectrum of velocity adopted from Ref. \citep{Wensink2012PNAS}. \red{Clearly the energy spectrum and dissipation spectrum peak at $k_{E}R\simeq 0.1$, corresponding to a spatial scale $\ell_E/R=10$, and $k_{\epsilon}R\simeq0.17$, corresponding to a spatial scale $\ell_\epsilon/R\simeq 6$, respectively. These two peak scales are comparable with the spatial size of the velocity field. Tentatively the scale $10R$ can be understood as a kind of influence scale for the present case, i.e. the energetic structure formed by the hydrodynamic interaction and constrained by the effective fluid viscosity.}}

\section{Streamline based intrinsic flow structure}
\begin{figure}
\centering
\includegraphics[width=1\linewidth,clip]{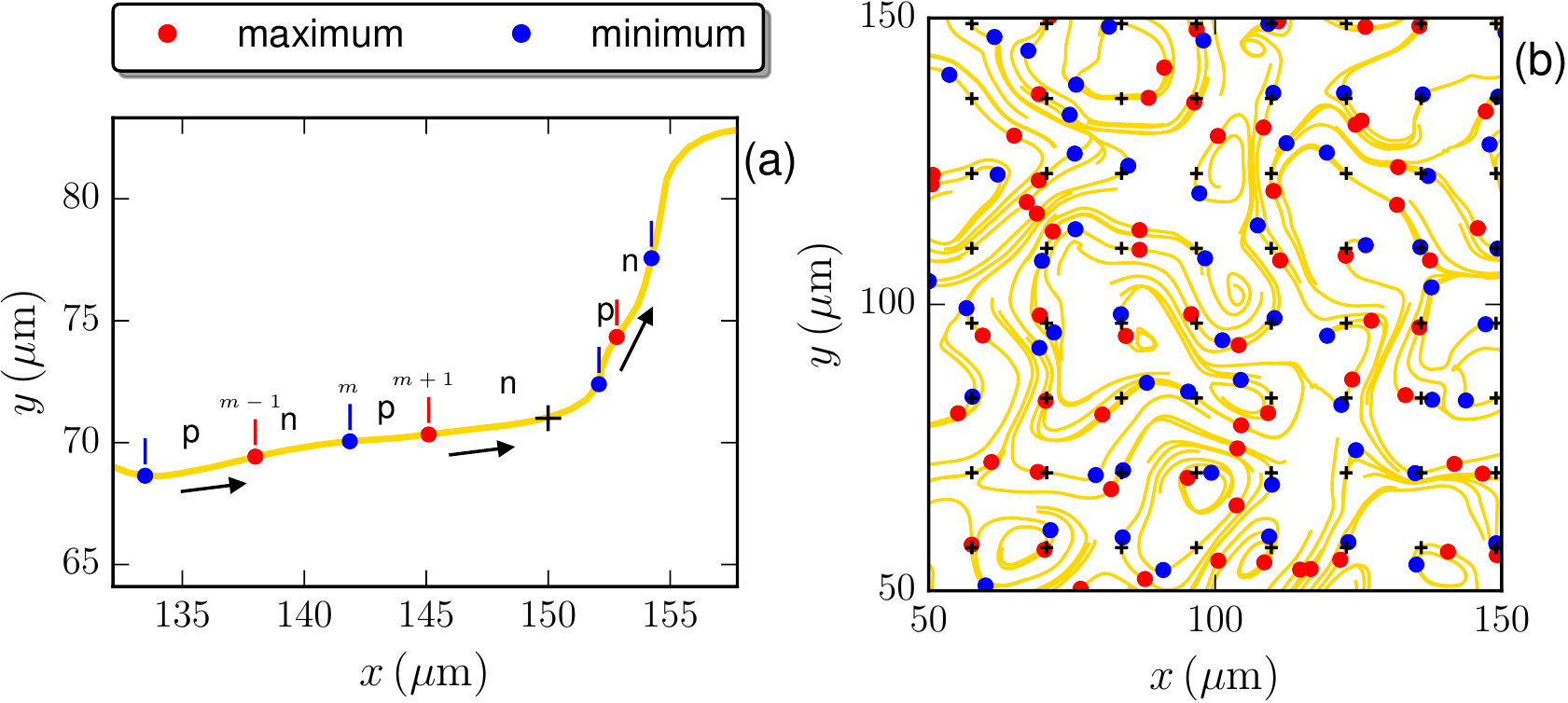}
  \caption{(Color online) a) Illustration of a  streamline with respect to a specified grid point (denoted by $+$). Along the streamline the local extremal points, either maximum (in red) or minimum (in blue), can be identified according to the velocity magnitude $u$. The given streamline is then divided into streamline segments, which are defined as the part confined by two adjacent local extremal points. According to the sign of $\Delta u$ along the velocity direction, the streamline segment can be positive (p) if $\Delta u>0$ or negative (n) if $\Delta u<0$. b) With respect to selected grid points (denoted by $+$), the corresponding streamline segment structure extracted from a snapshot experimental data with local maximal points and local minimal points. }\label{fig:Illustration}
\end{figure}

As shown in Fig.~\ref{fig:Illustration} (a) from a specific spatial point denoted by $+$, the streamline passing though is uniquely defined. Compared with other descriptions, the streamline is favorable to quantify the flow field because of the independence of the coordinate systems. Along the streamline the local extremal points, either maximum (in red) or minimum (in blue), can be identified according to the velocity magnitude $u$, which is calculated via the bilinear interpolation of each component. To ensure good resolution, numerically the spatial marching size along the streamline need to a fractal of the grid size, e.g. $0.02$. From the comparison of different algorithms, an ideal and robust criterion to find extrema is based on the interpolated velocity gradient $\nabla\vec{u}$, i.e. $\vec{n}\cdot\nabla\vec{u}\cdot\vec{n}=0$, where $\vec{n}=\vec{u}/u$ is unit velocity direction vector.

The streamline segment with respect to the given spatial point $+$ is defined as the part of the streamline containing $+$ and bounded by the two adjacent extremal points~\citep{Wang2010JFM}. Obviously within a streamline segment the velocity magnitude
changes monotonously. Along the velocity vector direction from one extremal with the velocity magnitude $u_s$ to another extremal point with the velocity magnitude $u_e$, $\Delta u=u_e-u_s$ and $\ell$, which is defined here as twice the curve length in between the adjacent extremal points, are the characteristic parameters of a streamline segment. Depending on the sign of $\Delta u$, the segment can be positive (p) if $\Delta u>0$ or negative (n) if $\Delta u<0$. As illustration, the streamline segments of some selected special points (denoted by $+$) are shown in Fig.~\ref{fig:Illustration} (b). An important issue to address here is that to make statistical results unbiased, the sampling special points, \red{through each a streamline segment passes}, need to be homogeneously distributed
~\citep{Wang2010JFM}.

The conventional structure function mixes information from different flow structures because the length scale $r$ is treated as an independent input \citep{Davidson2005PRL,Huang2013PRE,Schmitt2016Book}. It would be more serious  if an energetic structure presents, such as the observed large-structure here around $\sim10R$. It has been shown elsewhere that the classical structure function analysis is dominated by this structure \cite{Qiu2016PRE}. The scaling behavior predicted by the Fourier analysis is thus  absent in physical space \cite{Wensink2012PNAS}.  A more detail of scale dependent analysis can be found in Refs. \cite{Qiu2016PRE,Schmitt2016Book}.

The streamline segment concept is inspired by the fact that the turbulent vector field can naturally be described by the streamline, which is the coordinate system independent. The length scale of the streamline segment is determined by the extremal points of the velocity magnitude, but not an arbitrary input. Therefore, the streamline-based structure is advantageous in field description and scaling analysis to avoid scale mixing. Interesting examples of streamline segment analysis include the turbulent velocity field~\citep{Wang2010JFM}, the turbulent vorticity field~\citep{Wang2012PoF} and turbulent flame structure~\citep{Chakraborty2014PRE,Wang2013PCI}. Moreover, it needs to mention that because the streamline segment structure is defined based on the instantaneous flow field, thus the analysis is Eulerian but not Lagrangian.

\section{Results}

\subsection{Probability density function of the characteristic parameters}
\begin{figure}
\centering
\includegraphics[width=1\linewidth,clip]{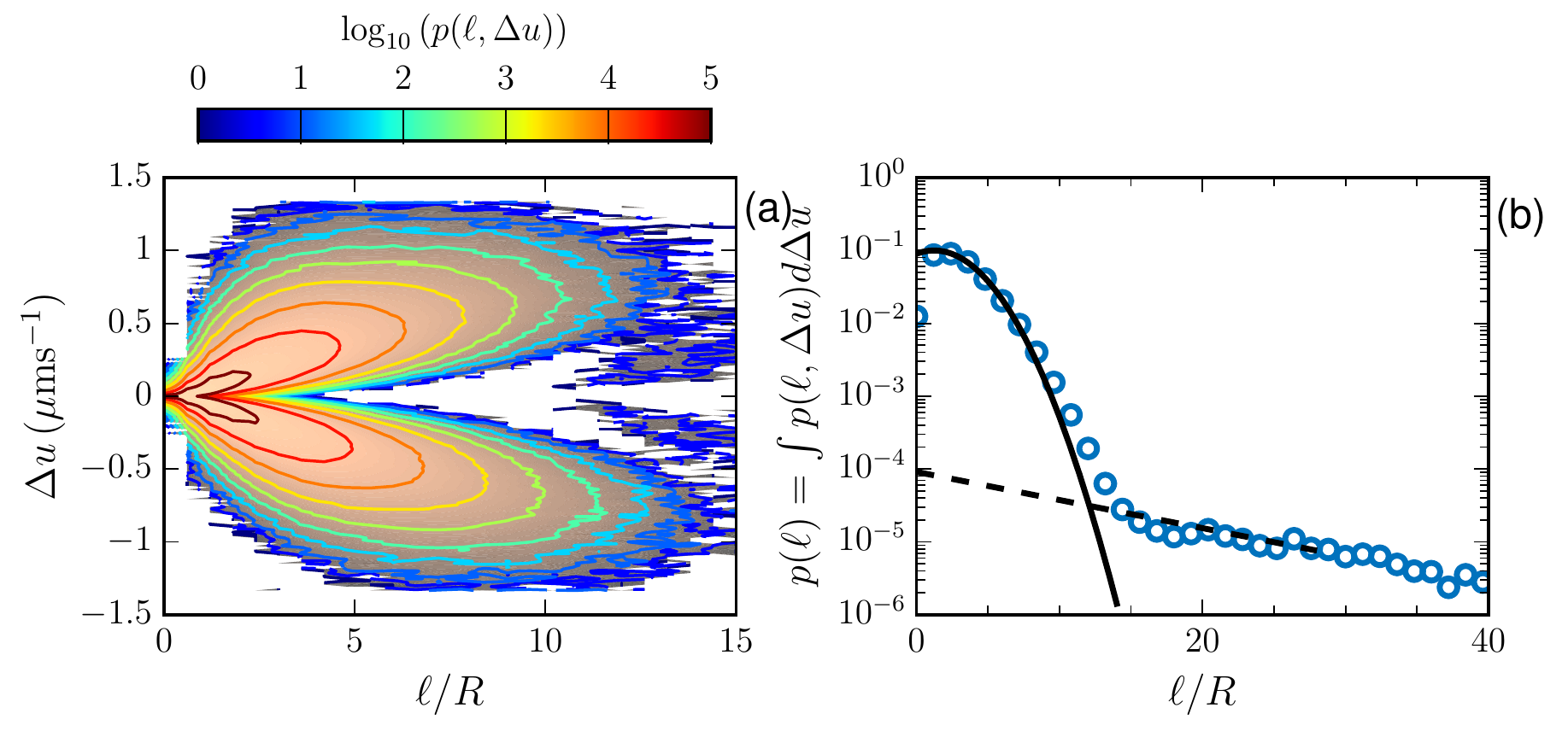}
  \caption{(Color online) a) Experimental joint-pdf $p(\ell,\Delta  u)$. 
  For display convenience, $p(\ell,\Delta  u)$ is measured at the logarithmic scale. \red{For a better display, the scale is in the range $0\le \ell/R\le 15$.} b) The corresponding marginal pdf $p(\ell)$. For comparison, normal and exponential  distributions are  illustrated as solid dashed  lines.} \label{fig:JPDF}
\end{figure}

The above mentioned streamline segment based analysis is applied to all snapshots. Totally, 9,092,689 segments are detected, which ensures a good statistics below.
Figure \ref{fig:JPDF}\,a) shows in the velocity field the joint probability density function (pdf) $p(\ell,\Delta u)$ between the two characteristic parameters, i.e. $\ell$ and $\Delta u$, for the streamline segments with respect to all the spatial points. The most noticeable feature is the asymmetry between the branch with positive $\Delta  u$ and the branch with negative $\Delta u $, i.e. on average the length of negative segments are larger than that of the positive segments, whereas $\Delta u$ for positive and negative segments are almost the same. The existing results for 3D fluid turbulence show such asymmetry structure as well~\citep{Wang2010JFM}. However, due to the kinematic effect along the velocity direction the positive segments have increasing $u $ and thus are inclined to be stretched and on average $\ell$ is larger;  whereas for the negative segments along the velocity direction $ u$ decreases and thus the segments tend to be compressed to have smaller $\ell$.  The opposite asymmetry we see here in bacterial turbulence, i.e. on average the negative streamline segments have larger $\ell$ than that of the positive streamline segments, is either a consequence of the dimensionality reduction from $3$ to $2$ or the active nature of this dynamic system. It deserves a more detailed further study.

The length scale pdf, i.e., $p(\ell)=\int p(\ell,\Delta u) d \Delta u$, is shown in Fig.~\ref{fig:JPDF}\,(b). Clearly there are two different regimes separated at $\ell/R\simeq 16$. Below this scale the pdf is roughly Gaussian, while a clear exponential tail appears above this scale, which is a strong evidence of the different dominate mechanisms in different regimes. \red{Here $\ell/R\simeq 16$ is comparable with the aforementioned influence scale. The observed two regimes may be attributed to the different influences from the viscous effect. As in the previous mechanism analysis \citep{Wang2010JFM,Wang2006JFM}, under the action of perturbation from the random motion of turbulent eddies, the extremal points of streamaline segments are newly generated; meanwhile the molecular diffusion will smear away the extremal points. In the smaller $\ell$ range the diffusion mechanism dominates, while for larger $\ell$ perturbation is more important.}


\subsection{High-order statistics and multifractality}


From the calculated joint-pdf, a $\ell$-based $q$th-order intrinsic structure function is introduced as
\begin{equation}
M_q(\ell)=\int_{-\infty}^{+\infty} p(\ell,\Delta u) \vert \Delta u\vert^q  d \Delta u.
\label{SF}\end{equation}
In the conventional definition of the structure function, the length scale is an independent input. Thus the average operation in the structure function mixes different correlation regions \citep{Wang2006JFM}. Mathematically, the structure function acts as a filter with a weight function $W(k\ell) =1-\cos(2\pi k\ell)$, in which $k$ is the wavenumber and $\ell$ is the separation scale \citep{Davidson2005PRL,Huang2011PRE,Schmitt2016Book}. It thus leads to the statistics at different wavenumber  $k$ mixed, resulting in the so-called infrared and ultraviolet effects, respectively for large-scale and small-scale contaminations \citep{Huang2013PRE}. In contrast, the length scale $\ell$ in equation~(\ref{SF}) is the segment length, which is determined by the intrinsic flow structure rather than an independent input. Such definition is in better agreement with the flow physics that scale is flow structure related, and at the same time helps to annihilate the strong mixing of different correlation regions. Therefore much improved results has been obtained in analyzing the Lagrangian and 2D turbulence data \citep{Wang2015JSTAT}.

\begin{figure}
\centering
\includegraphics[width=1\linewidth,clip]{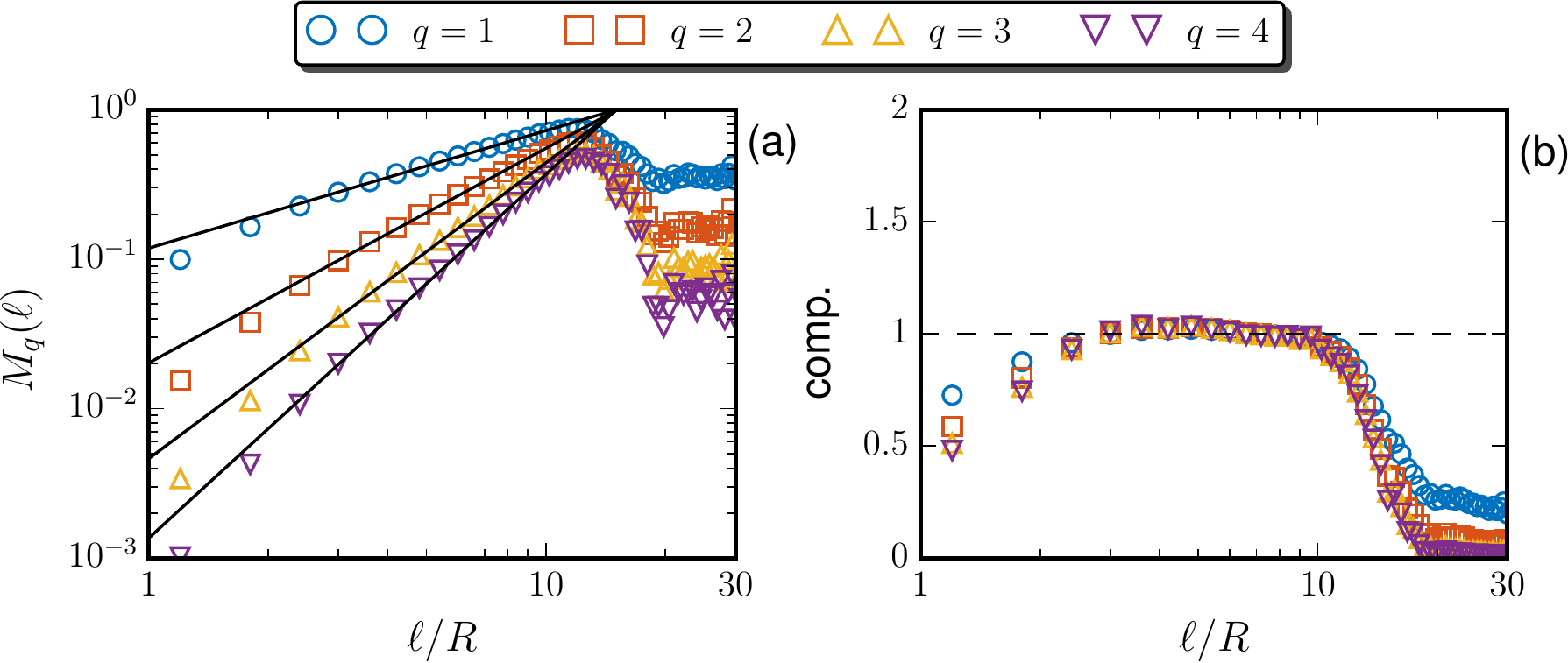}
  \caption{(Color online)
a) High-order intrinsic  structure function $M_q(\ell)$ for $q$ from 1 to 4. The Power-law behavior is observed in the range $2\le \ell/R\le 10$, corresponding to a wavenumber range $0.1\le kR\le 0.5$. The solid line indicates a power-law fitting in this scaling range via the least square fitting algorithm. b) The corresponding compensated curve using the fitted parameters to emphasize the observed power-law behavior. }\label{fig:Moments}
\end{figure}

We expect a scaling behavior of $M_q(\ell)$, e.g., $M_q(\ell)\sim \ell^{\zeta(q)}$, particularly at the separation scale $\ell/R\le 15$ with convergent statistics. The measured intrinsic structure functions $M_q(\ell)$ for $q=1, 2,3$ and $4$ are shown in Figure \ref{fig:Moments}\,a). The power-law behavior is observed in the range $2\le \ell/R\le 10$, corresponding to a wavenumber range $0.1\le kR\le 0.5$, which agrees well with the scaling range detected by the Hilbert-Huang transform in the wavenumber domain \citep{Qiu2016PRE}.  Figure \ref{fig:Moments}\,b) shows the corresponding compensated curve using the fitted parameter in a semi-log plot. A clear plateau confirms the existence of the scaling behavior.

\begin{figure}
\centering
\includegraphics[width=1\linewidth,clip]{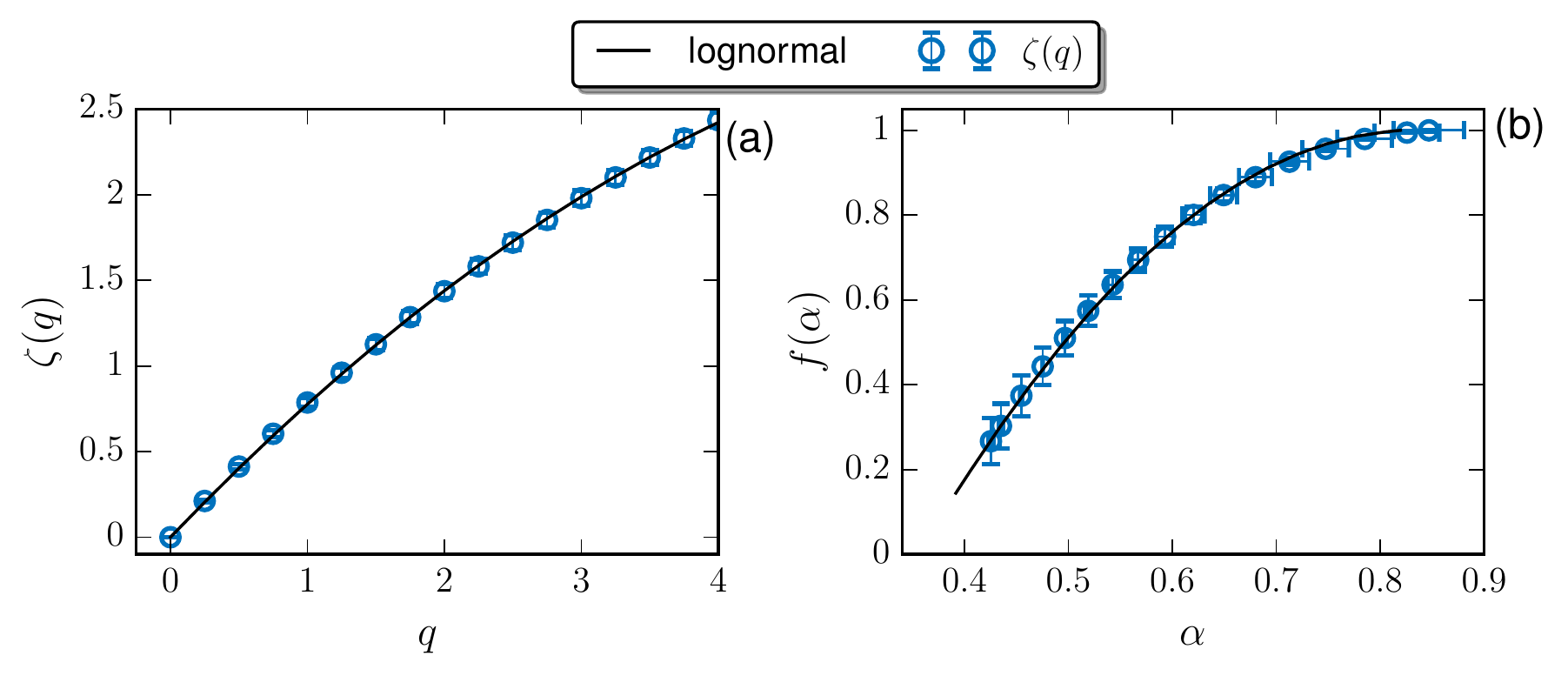}
  \caption{(Color online) a) Scaling exponent $\zeta(q)$ in the range $2\le \ell/R\le 10$ for $0\le q\le 4$. A lognormal formula with a Hurst number $H=0.76\pm0.01$ and  an intermittency parameter $\mu=0.20\pm0.01$ is shown as a solid line.  b) The corresponding singularity spectrum $f(\alpha)$.  The errorbar indicates a $95\%$ confidence interval. }\label{fig:Scaling}
\end{figure}

Figure \ref{fig:Scaling}\,a) displays scaling exponent $\zeta(q)$ extracted from the experimental data, where the errorbar indicates a $95\%$ fitting confidence interval. The convex shape implies that intermittency exists in this active dynamic system \citep{Qiu2016PRE}. Moreover the intermittency intensity can be quantified via the following lognormal formula,
\begin{equation}
\zeta(q)=qH-\frac{\mu}{2}\left(q^2H^2-qH \right), \label{eq:lognormal}
\end{equation}
where $H$ is the Hurst number and $\mu$ is the intermittency index, respectively. Mathematically $\mu$ characterizes how much the measured $\zeta(q)$ deviates from a linear relation $qH$. In other words, a larger $\mu$ is, more intermittent the process is.  The fitted results read $H=0.76\pm0.01$ and $\mu=0.20\pm0.01$.
Recently, the same level parameter $\mu=0.26\pm0.01$ has been reported by \citet{Qiu2016PRE} via a different methodology.
To our knowledge, the intermittency parameter $\mu_E\simeq 0.20$ has been widely accepted for the three-dimensional hydrodynamic turbulence \citep{Frisch1995}.  It is surprisingly interesting that different turbulent systems may assume the similar intermittent behavior. Even in the very low Reynolds number bacterial flow, intermittency, one of the most important turbulent features, still exists due to the strong nonlinear interaction between the background flow and the self-propulsion of the living organisms. The present result indicates that very different turbulent flows can still be similarly intermittent. 

\section{Discussion}
\subsection{Singularity spectrum}
To emphasize multifractality, a singularity spectrum is introduced via the following Legendre transform,
\begin{equation}
\alpha(q)=\frac{d \zeta(q)}{d q},\, f(\alpha)=\min_q\left\{ q\alpha -\zeta(q)+1 \right\},
\end{equation}
where $\alpha$ is a generalized Hurst number, and $f(\alpha)$ is the singularity spectrum. The broader variation $\alpha$ and $f(\alpha)$ implies a stronger multifractality of the process. Figure \ref{fig:Scaling}\,(b) shows the measured $f(\alpha)$ versus $\alpha$ with the $95\%$ confidence interval errorbars. Moreover, the measured singularity spectrum is well reproduced by the lognormal formula with the experimental Hurst number and intermittency parameter.

\subsection{Cascade direction}
\citet{Wensink2012PNAS} proposed the following continuum model of the bacterial turbulence
\begin{widetext}
\begin{equation}\label{eq:CT}
\partial_t \mathbf{u}+ \lambda_0\mathbf{u} \cdot\nabla \mathbf{u}=-\nabla p +\Gamma_0\nabla^2\mathbf{u}-\Gamma_2(\nabla^2)^2\mathbf{u}+\lambda_1 \nabla
\mathbf{u} ^2-(\varpi+ \chi\vert \mathbf{u}\vert^2) \mathbf{u},
\end{equation}
\end{widetext}
where $p$ denotes pressure; $\lambda_0>1$ and $\lambda_1>0$ are used for the pusher-swimmers as in this study; $(\varpi,\chi)$ corresponds to a quartic Landau-type velocity potential; $(\Gamma_0,\Gamma_2)$ provides the description of the self-sustained mesoscale turbulence in incompressible active flow, e.g., $\Gamma_0<0$ and $\Gamma_2>0$, resulting in a turbulent state \citep{Wensink2012PNAS}. As discussed in Ref.\,\citep{Qiu2016PRE}, the observed intermittency correction might be triggered by the last two nonlinear terms.
However, it is difficult to apply the above equation to the experiment data because of the intractability in determining the listed parameters. Alternative a two-dimensional Ekman-Navier-Stokes equation is considered as a first-order approximation of the governing equation, which is written as
\begin{equation}\label{eq:NSE}
\partial_t \mathbf{u}+\mathbf{u} \cdot\nabla \mathbf{u}=-\nabla p +\nu \nabla^2
\mathbf{u} -\xi \mathbf{u}+\mathbf{f}_{u},
\end{equation}
where $\xi$ stands for the Enkman friction coefficient, and $\mathbf{f}_{u}$ is the external forcing to inject the energy and enstrophy to the system~\citep{Boffetta2012ARFM}. The two-dimensional kinetic energy  and enstrophy fluxes can be derived via a filter-space technique \citep{Germano1992JFM,Ni2014PoF,Chen2003PRL,Boffetta2010PRE,Liao2014PoF,Zhou2015JFM}. Using the Gaussian filter $G^r=\sqrt{6/\pi}\exp(-6r^2)$ for instance, where $r$ is a coarse-grained scale, the filtered field is defined as
\begin{equation}
f^r(x)=\int_{\vert x'\vert \le r} G^r(x')f(x+x')\upd x'.
\end{equation}
It then yields
\begin{equation}
\Pi^E(r)=-\sum_{i,j=1,2}\left( \left( u_iu_j \right)^{r}-u_i^r u_j^r  \right)\frac{\partial u_i^r}{\partial x_j}
\end{equation}
for the energy flux and
\begin{equation}
\Pi^\Omega(r)=-\sum_{i=1,2}\left( \left( u_i\omega \right)^{r}-u_i^r \omega^r  \right)\frac{\partial \omega^r}{\partial x_i}
\end{equation}
for the enstrophy flux. A negative $\Pi^E(r)$ indicates an energy transferred from scale $<r$ to scale $>r$, and vice versa. This technique has been proved to be efficient even for analyzing the poorly resolved velocity field~\citep{Ni2014PoF}.

\begin{figure}
\centering
\includegraphics[width=0.85\linewidth,clip]{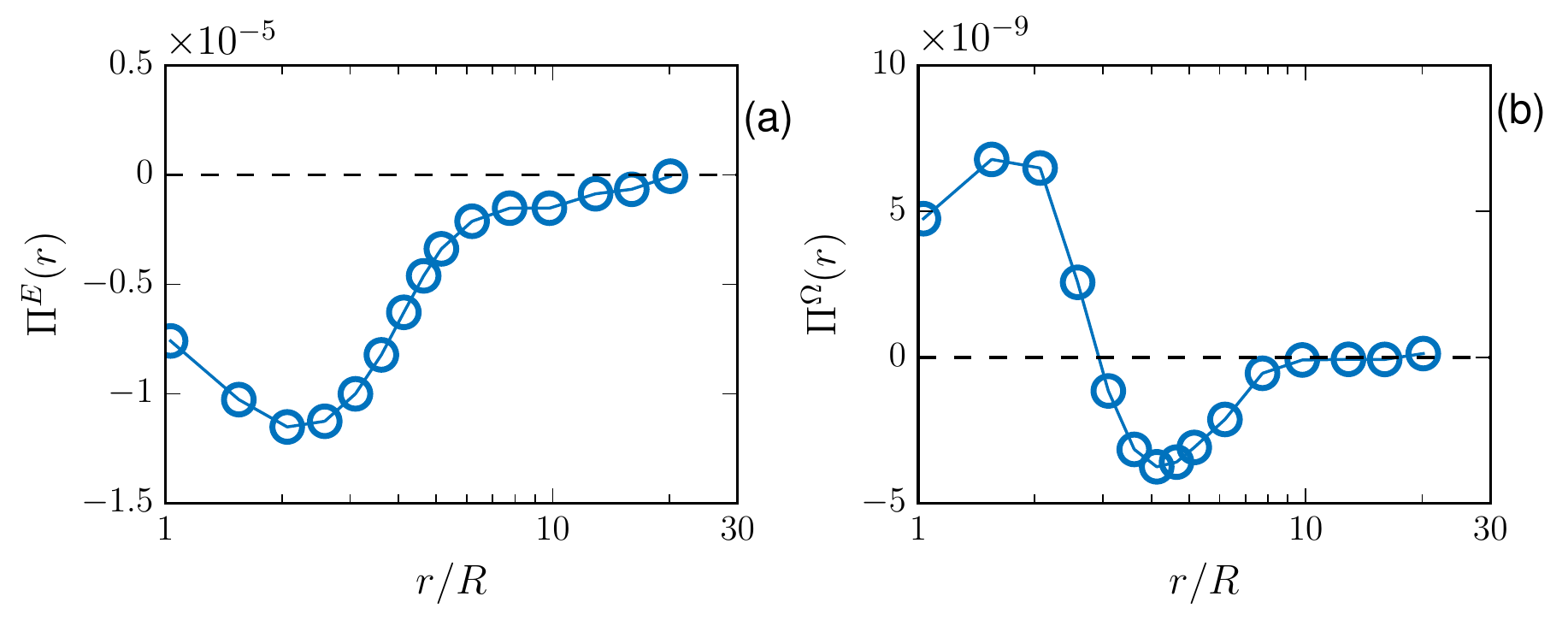}
  \caption{(Color online) Scale-to-scale flux for a) energy $\Pi^E(r)$ and b) enstrophy $\Pi^{\Omega}(r)$. A negative value indicates an inverse flux transferring from small to large scales.}\label{fig:Flux}
\end{figure}

Figure \ref{fig:Flux} (a) and (b) show the measured scale-to-scale the energy flux $\Pi^E(r)$ and enstrophy $\Pi^{\Omega}(r)$, respectively. The former one indicates an inverse energy cascade up to at least scale $r/R=20$. The latter one shows more complexity, e.g. a forward enstrophy cascade when $r/R\le 3$, and then an inverse cascade when  $3<r/R\le 10$. Note that the contribution from the additional nonlinear
interactions, i.e. the last two terms in Eq.~\eqref{eq:CT}, has been ignored, which, however, could be important for the enstrophy cascade since the vorticity is the first-order spatial derivative of the Eulerian velocity.
A following cascade picture can be postulated. The kinetic energy is injected into the system at the length scale of the bacterial body size $r/R=1$. It is then transferred up to large scales to generate large-scale motions. The fluid viscosity plays as an important role in this special inverse cascade at scales below $\ell_{\nu}$, which may physically act as an energy barrier to block the energy transfer toward larger scales.

\subsection{Lognormal statistics}
\begin{figure}
\centering
\includegraphics[width=0.85\linewidth,clip]{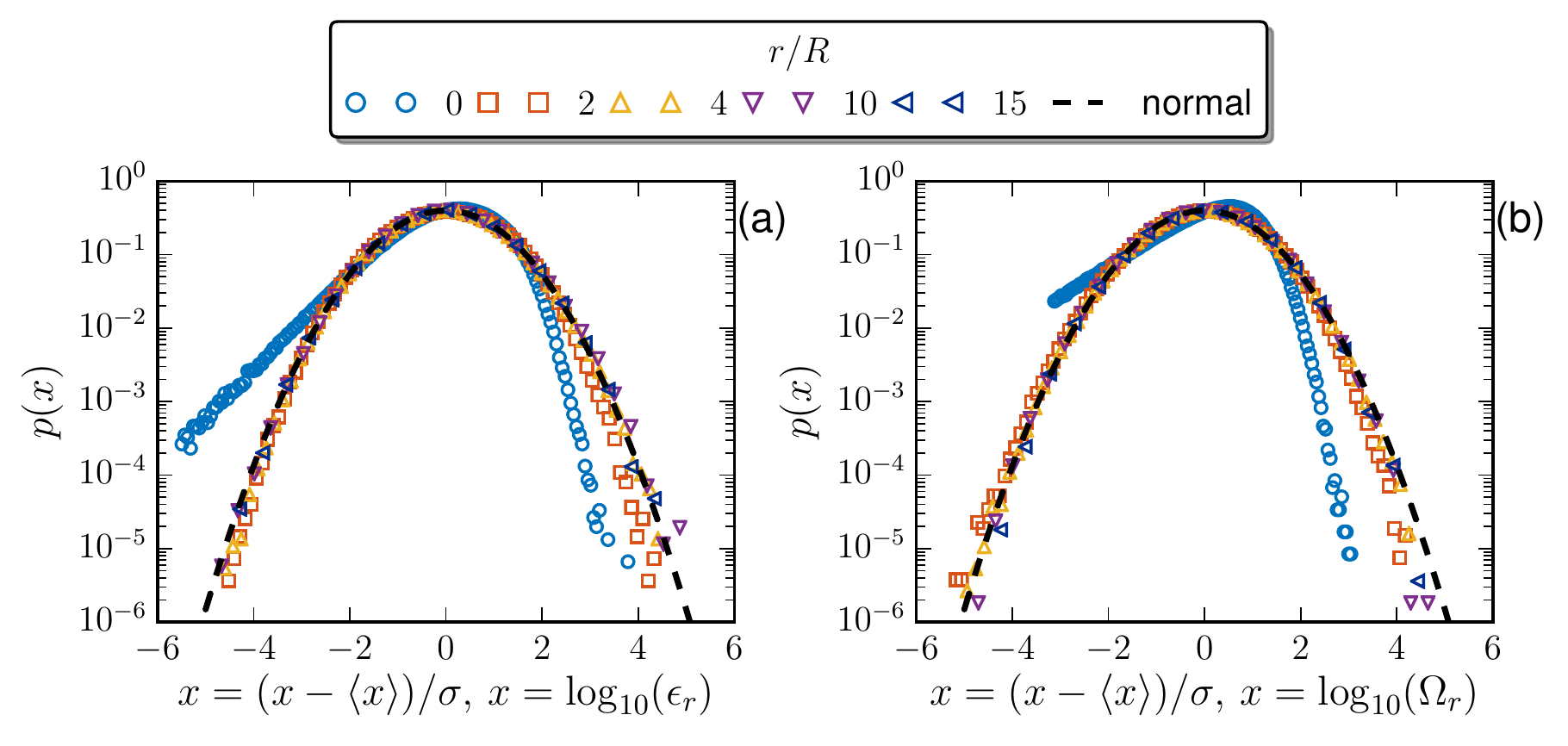}
  \caption{(Color online) Pdf of (a) the coarse-grained energy dissipation $\epsilon_r$; (b) the coarse-grained enstrophy $\Omega_r$. The lognormal formula is also shown (dashed line) for comparison.}\label{fig:CGpdf}
\end{figure}

The lognormal formula equation~(\ref{eq:lognormal}) is first introduced by Kolmogorov in his famous refined similarity hypothesis theory with $H=1/3$, where the intermittency property of the energy dissipation field ($\epsilon=\nu/2(\partial_i u_j+\partial_j u_i)^2$) is considered \citep{Kolmogorov1962}. He assumed a lognormal distribution of the coarse-grained energy dissipation, which is defined in a two-dimensional field as,
\begin{equation}
\epsilon_{r}(x)=\frac{1}{\pi r^2}\int_{\vert x'\vert\le r}\epsilon(x+x')\upd x',
\end{equation}
where $r$ is a coarse-grained scale. The same operation can be applied to the enstrophy $\Omega=\omega^2$.
\red{Note the fact that the velocity field in this active turbulence is smooth. The measured energy dissipation and esntrophy is thus less  influenced by measurement noise or spatial resolution.}
Figure \ref{fig:CGpdf} shows a test of the lognormal assumption at various scales for energy dissipation (a) and enstrophy (b). The results here confirms the validation of the lognormal assumption, which therefore is reasonable in this active dynamic system as well.

\section{Conclusions}
In summary, we proposed streamline segment analysis to extract multiscale information based on the intrinsic flow structure of the two-dimensional bacterial flow. The joint-pdf of measured $\ell$ and $\Delta u$ displays an asymmetric pattern, which can be linked with the special features, e.g. the cascade process, of this two-dimensional active dynamic system. The marginal distribution of the scale $\ell$ shows two different regimes, which are separated at $\ell/R\simeq 16$.  This two regime behavior is related to the change of the the viscous influence with the scale. Compared with the conventional structure function, the $\ell$-based definition captures the flow structure in a more natural way, showing evidently a nearly one decade power-law range. The scaling exponent $\zeta(q)$ and singularity spectrum can be described nicely by a lognormal formula with the intermittency parameter $\mu=0.20$, which agrees closely with the result for three-dimensional hydrodynamic turbulence. This observed intermittency universality is important to understand the turbulence physics.

Moreover, the direction of the energy and enstrophy cascade is measured via the filter-space technique. The experiment result confirms an inverse energy cascade as expected for this active dynamic system. Concerning the enstrophy cascade, it is more complex than the energy case. The present results suggest a forward cascade when $r/R\le 3$, while an inverse one when $r/R>3$. Additionally, the lognormal statistics is verified for both the energy dissipation field and the enstrophy field.

\red{Finally, we provide a comment on the results obtained in this work. The dynamical property of the bacterial turbulence may depend on several facts, for instance the type of bacteria, concentration, temperature, etc. Therefore, the measured intermittency parameter $\mu$ and the Hurst number $H$ may also show such dependence.  A systematical parametric study should be conducted to check whether the two-dimensional bacterial turbulence share the same intermittency parameter as three-dimensional hydrodynamical turbulence or not. Moreover, a more detailed study of the inverse energy cascade in this active dynamical system will enrich our understanding of not only the bacterial turbulence, but also the high Reynolds number fluid turbulence, where the inverse energy cascade exists.}

\begin{acknowledgments}
This work is sponsored by the National Natural Science Foundation of China (under Grant Nos. 11332006 and 91441116), and  partially by the Sino-French (NSFC-CNRS) joint research project (No. 11611130099, NSFC China, and PRC 2016-2018 LATUMAR ``Turbulence lagrangienne: \'etudes num\'eriques et applications environnementales marines",  CNRS, France).  Y.H. is also supported by the Fundamental Research Funds for the Central Universities (Grant No. 20720150075). We thank Prof. R.~E. Goldstein for providing the experimental data, which can be found at \footnote{See {http://damtp.cam.ac.uk/user/gold/datarequests.html}}.
A source package for analyzing the streamline segment structure is available at \footnote{See {https://github.com/lanlankai}}.
\end{acknowledgments}


\begin{thebibliography}{42}%
\makeatletter
\providecommand \@ifxundefined [1]{%
 \@ifx{#1\undefined}
}%
\providecommand \@ifnum [1]{%
 \ifnum #1\expandafter \@firstoftwo
 \else \expandafter \@secondoftwo
 \fi
}%
\providecommand \@ifx [1]{%
 \ifx #1\expandafter \@firstoftwo
 \else \expandafter \@secondoftwo
 \fi
}%
\providecommand \natexlab [1]{#1}%
\providecommand \enquote  [1]{``#1''}%
\providecommand \bibnamefont  [1]{#1}%
\providecommand \bibfnamefont [1]{#1}%
\providecommand \citenamefont [1]{#1}%
\providecommand \href@noop [0]{\@secondoftwo}%
\providecommand \href [0]{\begingroup \@sanitize@url \@href}%
\providecommand \@href[1]{\@@startlink{#1}\@@href}%
\providecommand \@@href[1]{\endgroup#1\@@endlink}%
\providecommand \@sanitize@url [0]{\catcode `\\12\catcode `\$12\catcode
  `\&12\catcode `\#12\catcode `\^12\catcode `\_12\catcode `\%12\relax}%
\providecommand \@@startlink[1]{}%
\providecommand \@@endlink[0]{}%
\providecommand \url  [0]{\begingroup\@sanitize@url \@url }%
\providecommand \@url [1]{\endgroup\@href {#1}{\urlprefix }}%
\providecommand \urlprefix  [0]{URL }%
\providecommand \Eprint [0]{\href }%
\providecommand \doibase [0]{http://dx.doi.org/}%
\providecommand \selectlanguage [0]{\@gobble}%
\providecommand \bibinfo  [0]{\@secondoftwo}%
\providecommand \bibfield  [0]{\@secondoftwo}%
\providecommand \translation [1]{[#1]}%
\providecommand \BibitemOpen [0]{}%
\providecommand \bibitemStop [0]{}%
\providecommand \bibitemNoStop [0]{.\EOS\space}%
\providecommand \EOS [0]{\spacefactor3000\relax}%
\providecommand \BibitemShut  [1]{\csname bibitem#1\endcsname}%
\let\auto@bib@innerbib\@empty
\bibitem [{\citenamefont {Frisch}(1995)}]{Frisch1995}%
  \BibitemOpen
  \bibfield  {author} {\bibinfo {author} {\bibfnamefont {U.}~\bibnamefont
  {Frisch}},\ }\href@noop {} {\emph {\bibinfo {title} {{Turbulence: the legacy
  of AN Kolmogorov}}}}\ (\bibinfo  {publisher} {Cambridge University Press},\
  \bibinfo {year} {1995})\BibitemShut {NoStop}%
\bibitem [{\citenamefont {Wensink}\ \emph {et~al.}(2012)\citenamefont
  {Wensink}, \citenamefont {Dunkel}, \citenamefont {Heidenreich}, \citenamefont
  {Drescher}, \citenamefont {Goldstein}, \citenamefont {L{\"o}wen},\ and\
  \citenamefont {Yeomans}}]{Wensink2012PNAS}%
  \BibitemOpen
  \bibfield  {author} {\bibinfo {author} {\bibfnamefont {H.}~\bibnamefont
  {Wensink}}, \bibinfo {author} {\bibfnamefont {J.}~\bibnamefont {Dunkel}},
  \bibinfo {author} {\bibfnamefont {S.}~\bibnamefont {Heidenreich}}, \bibinfo
  {author} {\bibfnamefont {K.}~\bibnamefont {Drescher}}, \bibinfo {author}
  {\bibfnamefont {R.}~\bibnamefont {Goldstein}}, \bibinfo {author}
  {\bibfnamefont {H.}~\bibnamefont {L{\"o}wen}}, \ and\ \bibinfo {author}
  {\bibfnamefont {J.}~\bibnamefont {Yeomans}},\ }\href@noop {} {\bibfield
  {journal} {\bibinfo  {journal} {PNAS}\ }\textbf {\bibinfo {volume} {109}},\
  \bibinfo {pages} {14308} (\bibinfo {year} {2012})}\BibitemShut {NoStop}%
\bibitem [{\citenamefont {Davidson}\ and\ \citenamefont
  {Pearson}(2005)}]{Davidson2005PRL}%
  \BibitemOpen
  \bibfield  {author} {\bibinfo {author} {\bibfnamefont {P.~A.}\ \bibnamefont
  {Davidson}}\ and\ \bibinfo {author} {\bibfnamefont {B.~R.}\ \bibnamefont
  {Pearson}},\ }\href@noop {} {\bibfield  {journal} {\bibinfo  {journal} {Phys.
  Rev. Lett.}\ }\textbf {\bibinfo {volume} {95}},\ \bibinfo {pages} {214501}
  (\bibinfo {year} {2005})}\BibitemShut {NoStop}%
\bibitem [{\citenamefont {Huang}\ \emph {et~al.}(2011)\citenamefont {Huang},
  \citenamefont {Schmitt}, \citenamefont {Hermand}, \citenamefont {Gagne},
  \citenamefont {Lu},\ and\ \citenamefont {Liu}}]{Huang2011PRE}%
  \BibitemOpen
  \bibfield  {author} {\bibinfo {author} {\bibfnamefont {Y.}~\bibnamefont
  {Huang}}, \bibinfo {author} {\bibfnamefont {F.~G.}~\bibnamefont {Schmitt}},
  \bibinfo {author} {\bibfnamefont {J.-P.}\ \bibnamefont {Hermand}}, \bibinfo
  {author} {\bibfnamefont {Y.}~\bibnamefont {Gagne}}, \bibinfo {author}
  {\bibfnamefont {Z.}~\bibnamefont {Lu}}, \ and\ \bibinfo {author}
  {\bibfnamefont {Y.}~\bibnamefont {Liu}},\ }\href@noop {} {\bibfield
  {journal} {\bibinfo  {journal} {Phys. Rev. E}\ }\textbf {\bibinfo {volume}
  {84}},\ \bibinfo {pages} {016208} (\bibinfo {year} {2011})}\BibitemShut
  {NoStop}%
\bibitem [{\citenamefont {Schmitt}\ and\ \citenamefont
  {Huang}(2016)}]{Schmitt2016Book}%
  \BibitemOpen
  \bibfield  {author} {\bibinfo {author} {\bibfnamefont {F.}~\bibnamefont
  {Schmitt}}\ and\ \bibinfo {author} {\bibfnamefont {Y.}~\bibnamefont
  {Huang}},\ }\href@noop {} {\emph {\bibinfo {title} {Stochastic Analysis of
  Scaling Time Series: From Turbulence Theory to Applications}}}\ (\bibinfo
  {publisher} {Cambridge Univ Press},\ \bibinfo {year} {2016})\BibitemShut
  {NoStop}%
\bibitem [{\citenamefont {Peng}\ \emph {et~al.}(1994)\citenamefont {Peng},
  \citenamefont {Buldyrev}, \citenamefont {Havlin}, \citenamefont {Simons},
  \citenamefont {Stanley},\ and\ \citenamefont {Goldberger}}]{Peng1994PRE}%
  \BibitemOpen
  \bibfield  {author} {\bibinfo {author} {\bibfnamefont {C.~K.}~\bibnamefont
  {Peng}}, \bibinfo {author} {\bibfnamefont {S.~V.}~\bibnamefont {Buldyrev}},
  \bibinfo {author} {\bibfnamefont {S.}~\bibnamefont {Havlin}}, \bibinfo
  {author} {\bibfnamefont {M.}~\bibnamefont {Simons}}, \bibinfo {author}
  {\bibfnamefont {H.~E.}~\bibnamefont {Stanley}}, \ and\ \bibinfo {author}
  {\bibfnamefont {A.~L.}~\bibnamefont {Goldberger}},\ }\href@noop {} {\bibfield
  {journal} {\bibinfo  {journal} {Phys. Rev. E}\ }\textbf {\bibinfo {volume}
  {49}},\ \bibinfo {pages} {1685} (\bibinfo {year} {1994})}\BibitemShut
  {NoStop}%
\bibitem [{\citenamefont {Huang}\ \emph {et~al.}(1998)\citenamefont {Huang},
  \citenamefont {Shen}, \citenamefont {Long}, \citenamefont {Wu}, \citenamefont
  {Shih}, \citenamefont {Zheng}, \citenamefont {Yen}, \citenamefont {Tung},\
  and\ \citenamefont {Liu}}]{Huang1998EMD}%
  \BibitemOpen
  \bibfield  {author} {\bibinfo {author} {\bibfnamefont {N.}~\bibnamefont
  {Huang}}, \bibinfo {author} {\bibfnamefont {Z.}~\bibnamefont {Shen}},
  \bibinfo {author} {\bibfnamefont {S.}~\bibnamefont {Long}}, \bibinfo {author}
  {\bibfnamefont {M.}~\bibnamefont {Wu}}, \bibinfo {author} {\bibfnamefont
  {H.}~\bibnamefont {Shih}}, \bibinfo {author} {\bibfnamefont {Q.}~\bibnamefont
  {Zheng}}, \bibinfo {author} {\bibfnamefont {N.}~\bibnamefont {Yen}}, \bibinfo
  {author} {\bibfnamefont {C.}~\bibnamefont {Tung}}, \ and\ \bibinfo {author}
  {\bibfnamefont {H.}~\bibnamefont {Liu}},\ }\href@noop {} {\bibfield
  {journal} {\bibinfo  {journal} {Proc. R. Soc. London, Ser. A}\ }\textbf
  {\bibinfo {volume} {454}},\ \bibinfo {pages} {903} (\bibinfo {year}
  {1998})}\BibitemShut {NoStop}%
\bibitem [{\citenamefont {Huang}\ \emph {et~al.}(2008)\citenamefont {Huang},
  \citenamefont {Schmitt}, \citenamefont {Lu},\ and\ \citenamefont
  {Liu}}]{Huang2008EPL}%
  \BibitemOpen
  \bibfield  {author} {\bibinfo {author} {\bibfnamefont {Y.}~\bibnamefont
  {Huang}}, \bibinfo {author} {\bibfnamefont {F.}~\bibnamefont {Schmitt}},
  \bibinfo {author} {\bibfnamefont {Z.}~\bibnamefont {Lu}}, \ and\ \bibinfo
  {author} {\bibfnamefont {Y.}~\bibnamefont {Liu}},\ }\href@noop {} {\bibfield
  {journal} {\bibinfo  {journal} {Europhys. Lett.}\ }\textbf {\bibinfo {volume}
  {84}},\ \bibinfo {pages} {40010} (\bibinfo {year} {2008})}\BibitemShut
  {NoStop}%
\bibitem [{\citenamefont {Wang}\ and\ \citenamefont
  {Huang}(2015)}]{Wang2015JSTAT}%
  \BibitemOpen
  \bibfield  {author} {\bibinfo {author} {\bibfnamefont {L.}~\bibnamefont
  {Wang}}\ and\ \bibinfo {author} {\bibfnamefont {Y.}~\bibnamefont {Huang}},\
  }\href@noop {} {\bibfield  {journal} {\bibinfo  {journal} {J. Stat. Mech.
  Theor. Exp.}\ ,\ \bibinfo {pages} {P06018}} (\bibinfo {year}
  {2015})}\BibitemShut {NoStop}%
\bibitem [{\citenamefont {Thampi}\ \emph {et~al.}(2013)\citenamefont {Thampi},
  \citenamefont {Golestanian},\ and\ \citenamefont {Yeomans}}]{Thampi2013PRL}%
  \BibitemOpen
  \bibfield  {author} {\bibinfo {author} {\bibfnamefont {S.~P.}~\bibnamefont
  {Thampi}}, \bibinfo {author} {\bibfnamefont {R.}~\bibnamefont {Golestanian}},
  \ and\ \bibinfo {author} {\bibfnamefont {J.~M.}~\bibnamefont {Yeomans}},\
  }\href@noop {} {\bibfield  {journal} {\bibinfo  {journal} {Phys. Rev. Lett.}\
  }\textbf {\bibinfo {volume} {111}},\ \bibinfo {pages} {118101} (\bibinfo
  {year} {2013})}\BibitemShut {NoStop}%
\bibitem [{\citenamefont {Thampi}\ \emph {et~al.}(2014)\citenamefont {Thampi},
  \citenamefont {Golestanian},\ and\ \citenamefont {Yeomans}}]{Thampi2014PTRS}%
  \BibitemOpen
  \bibfield  {author} {\bibinfo {author} {\bibfnamefont {S.~P.}~\bibnamefont
  {Thampi}}, \bibinfo {author} {\bibfnamefont {R.}~\bibnamefont {Golestanian}},
  \ and\ \bibinfo {author} {\bibfnamefont {J.~M.}~\bibnamefont {Yeomans}},\
  }\href@noop {} {\bibfield  {journal} {\bibinfo  {journal} {Phil. Trans. R.
  Soc. A}\ }\textbf {\bibinfo {volume} {372}},\ \bibinfo {pages} {20130366}
  (\bibinfo {year} {2014})}\BibitemShut {NoStop}%
\bibitem [{\citenamefont {Fodor}\ \emph {et~al.}(2016)\citenamefont {Fodor},
  \citenamefont {Nardini}, \citenamefont {Cates}, \citenamefont {Tailleur},
  \citenamefont {Visco},\ and\ \citenamefont {van Wijland}}]{Fodor2016PRL}%
  \BibitemOpen
  \bibfield  {author} {\bibinfo {author} {\bibfnamefont {{\'E}.}~\bibnamefont
  {Fodor}}, \bibinfo {author} {\bibfnamefont {C.}~\bibnamefont {Nardini}},
  \bibinfo {author} {\bibfnamefont {M.~E.}~\bibnamefont {Cates}}, \bibinfo
  {author} {\bibfnamefont {J.}~\bibnamefont {Tailleur}}, \bibinfo {author}
  {\bibfnamefont {P.}~\bibnamefont {Visco}}, \ and\ \bibinfo {author}
  {\bibfnamefont {F.}~\bibnamefont {van Wijland}},\ }\href@noop {} {\bibfield
  {journal} {\bibinfo  {journal} {Phy. Rev. Lett.}\ }\textbf {\bibinfo {volume}
  {117}},\ \bibinfo {pages} {038103} (\bibinfo {year} {2016})}\BibitemShut
  {NoStop}%
\bibitem [{\citenamefont {Bratanov}\ \emph {et~al.}(2015)\citenamefont
  {Bratanov}, \citenamefont {Jenko},\ and\ \citenamefont
  {Frey}}]{Bratanov2015PNAS}%
  \BibitemOpen
  \bibfield  {author} {\bibinfo {author} {\bibfnamefont {V.}~\bibnamefont
  {Bratanov}}, \bibinfo {author} {\bibfnamefont {F.}~\bibnamefont {Jenko}}, \
  and\ \bibinfo {author} {\bibfnamefont {E.}~\bibnamefont {Frey}},\ }\href@noop
  {} {\bibfield  {journal} {\bibinfo  {journal} {PNAS}\ }\textbf {\bibinfo
  {volume} {112}},\ \bibinfo {pages} {15048} (\bibinfo {year}
  {2015})}\BibitemShut {NoStop}%
\bibitem [{\citenamefont {Wu}\ and\ \citenamefont
  {Libchaber}(2000)}]{Wu2000PRL}%
  \BibitemOpen
  \bibfield  {author} {\bibinfo {author} {\bibfnamefont {X.-L.}\ \bibnamefont
  {Wu}}\ and\ \bibinfo {author} {\bibfnamefont {A.}~\bibnamefont {Libchaber}},\
  }\href@noop {} {\bibfield  {journal} {\bibinfo  {journal} {Phys. Rev. Lett.}\
  }\textbf {\bibinfo {volume} {84}},\ \bibinfo {pages} {3017} (\bibinfo {year}
  {2000})}\BibitemShut {NoStop}%
\bibitem [{\citenamefont {Pooley}\ \emph {et~al.}(2007)\citenamefont {Pooley},
  \citenamefont {Alexander},\ and\ \citenamefont {Yeomans}}]{Pooley2007PRL}%
  \BibitemOpen
  \bibfield  {author} {\bibinfo {author} {\bibfnamefont {C.~M.}\ \bibnamefont
  {Pooley}}, \bibinfo {author} {\bibfnamefont {G.~P.}\ \bibnamefont
  {Alexander}}, \ and\ \bibinfo {author} {\bibfnamefont {J.~M.}\ \bibnamefont
  {Yeomans}},\ }\href@noop {} {\bibfield  {journal} {\bibinfo  {journal} {Phys.
  Rev. Lett.}\ }\textbf {\bibinfo {volume} {99}},\ \bibinfo {pages} {228103}
  (\bibinfo {year} {2007})}\BibitemShut {NoStop}%
\bibitem [{\citenamefont {Ishikawa}\ and\ \citenamefont
  {Pedley}(2008)}]{Ishikawa2008PRL}%
  \BibitemOpen
  \bibfield  {author} {\bibinfo {author} {\bibfnamefont {T.}~\bibnamefont
  {Ishikawa}}\ and\ \bibinfo {author} {\bibfnamefont {T.~J.}\ \bibnamefont
  {Pedley}},\ }\href@noop {} {\bibfield  {journal} {\bibinfo  {journal} {Phys.
  Rev. Lett.}\ }\textbf {\bibinfo {volume} {100}},\ \bibinfo {pages} {088103}
  (\bibinfo {year} {2008})}\BibitemShut {NoStop}%
\bibitem [{\citenamefont {Rushkin}\ \emph {et~al.}(2010)\citenamefont
  {Rushkin}, \citenamefont {Kantsler},\ and\ \citenamefont
  {Goldstein}}]{Rushkin2010PRL}%
  \BibitemOpen
  \bibfield  {author} {\bibinfo {author} {\bibfnamefont {I.}~\bibnamefont
  {Rushkin}}, \bibinfo {author} {\bibfnamefont {V.}~\bibnamefont {Kantsler}}, \
  and\ \bibinfo {author} {\bibfnamefont {R.~E.}~\bibnamefont {Goldstein}},\
  }\href@noop {} {\bibfield  {journal} {\bibinfo  {journal} {Phys. Rev. Lett.}\
  }\textbf {\bibinfo {volume} {105}},\ \bibinfo {pages} {188101} (\bibinfo
  {year} {2010})}\BibitemShut {NoStop}%
\bibitem [{\citenamefont {Ishikawa}\ \emph {et~al.}(2011)\citenamefont
  {Ishikawa}, \citenamefont {Yoshida}, \citenamefont {Ueno}, \citenamefont
  {Wiedeman}, \citenamefont {Imai},\ and\ \citenamefont
  {Yamaguchi}}]{Ishikawa2011PRL}%
  \BibitemOpen
  \bibfield  {author} {\bibinfo {author} {\bibfnamefont {T.}~\bibnamefont
  {Ishikawa}}, \bibinfo {author} {\bibfnamefont {N.}~\bibnamefont {Yoshida}},
  \bibinfo {author} {\bibfnamefont {H.}~\bibnamefont {Ueno}}, \bibinfo {author}
  {\bibfnamefont {M.}~\bibnamefont {Wiedeman}}, \bibinfo {author}
  {\bibfnamefont {Y.}~\bibnamefont {Imai}}, \ and\ \bibinfo {author}
  {\bibfnamefont {T.}~\bibnamefont {Yamaguchi}},\ }\href@noop {} {\bibfield
  {journal} {\bibinfo  {journal} {Phys. Rev. Lett.}\ }\textbf {\bibinfo
  {volume} {107}},\ \bibinfo {pages} {028102} (\bibinfo {year}
  {2011})}\BibitemShut {NoStop}%
\bibitem [{\citenamefont {Chen}\ \emph {et~al.}(2012)\citenamefont {Chen},
  \citenamefont {Dong}, \citenamefont {Be'er}, \citenamefont {Swinney},\ and\
  \citenamefont {Zhang}}]{Chen2012PRL}%
  \BibitemOpen
  \bibfield  {author} {\bibinfo {author} {\bibfnamefont {X.}~\bibnamefont
  {Chen}}, \bibinfo {author} {\bibfnamefont {X.}~\bibnamefont {Dong}}, \bibinfo
  {author} {\bibfnamefont {A.}~\bibnamefont {Be'er}}, \bibinfo {author}
  {\bibfnamefont {H.~L.}~\bibnamefont {Swinney}}, \ and\ \bibinfo {author}
  {\bibfnamefont {H.~P.}~\bibnamefont {Zhang}},\ }\href@noop {} {\bibfield
  {journal} {\bibinfo  {journal} {Phys. Rev. Lett.}\ }\textbf {\bibinfo
  {volume} {108}},\ \bibinfo {pages} {148101} (\bibinfo {year}
  {2012})}\BibitemShut {NoStop}%
\bibitem [{\citenamefont {Dunkel}\ \emph
  {et~al.}(2013{\natexlab{a}})\citenamefont {Dunkel}, \citenamefont
  {Heidenreich}, \citenamefont {Drescher}, \citenamefont {Wensink},
  \citenamefont {B{\"a}r},\ and\ \citenamefont {Goldstein}}]{Dunkel2013PRL}%
  \BibitemOpen
  \bibfield  {author} {\bibinfo {author} {\bibfnamefont {J.}~\bibnamefont
  {Dunkel}}, \bibinfo {author} {\bibfnamefont {S.}~\bibnamefont {Heidenreich}},
  \bibinfo {author} {\bibfnamefont {K.}~\bibnamefont {Drescher}}, \bibinfo
  {author} {\bibfnamefont {H.~H.}~\bibnamefont {Wensink}}, \bibinfo {author}
  {\bibfnamefont {M.}~\bibnamefont {B{\"a}r}}, \ and\ \bibinfo {author}
  {\bibfnamefont {R.~E.}~\bibnamefont {Goldstein}},\ }\href@noop {} {\bibfield
  {journal} {\bibinfo  {journal} {Phys. Rev. Lett.}\ }\textbf {\bibinfo
  {volume} {110}},\ \bibinfo {pages} {228102} (\bibinfo {year}
  {2013}{\natexlab{a}})}\BibitemShut {NoStop}%
\bibitem [{\citenamefont {Saintillan}\ and\ \citenamefont
  {Shelley}(2012)}]{Saintillan2012JRSI}%
  \BibitemOpen
  \bibfield  {author} {\bibinfo {author} {\bibfnamefont {D.}~\bibnamefont
  {Saintillan}}\ and\ \bibinfo {author} {\bibfnamefont {M.}~\bibnamefont
  {Shelley}},\ }\href@noop {} {\bibfield  {journal} {\bibinfo  {journal} {J. R.
  Soc. Interface}\ }\textbf {\bibinfo {volume} {9}},\ \bibinfo {pages} {571}
  (\bibinfo {year} {2012})}\BibitemShut {NoStop}%
\bibitem [{\citenamefont {Dunkel}\ \emph
  {et~al.}(2013{\natexlab{b}})\citenamefont {Dunkel}, \citenamefont
  {Heidenreich}, \citenamefont {B{\"a}r},\ and\ \citenamefont
  {Goldstein}}]{Dunkel2013NJP}%
  \BibitemOpen
  \bibfield  {author} {\bibinfo {author} {\bibfnamefont {J.}~\bibnamefont
  {Dunkel}}, \bibinfo {author} {\bibfnamefont {S.}~\bibnamefont {Heidenreich}},
  \bibinfo {author} {\bibfnamefont {M.}~\bibnamefont {B{\"a}r}}, \ and\
  \bibinfo {author} {\bibfnamefont {R.}~\bibnamefont {Goldstein}},\ }\href@noop
  {} {\bibfield  {journal} {\bibinfo  {journal} {New J. Phys.}\ }\textbf
  {\bibinfo {volume} {15}},\ \bibinfo {pages} {045016} (\bibinfo {year}
  {2013}{\natexlab{b}})}\BibitemShut {NoStop}%
\bibitem [{\citenamefont {Gro\ss{}mann}\ \emph {et~al.}(2014)\citenamefont
  {Gro\ss{}mann}, \citenamefont {Romanczuk}, \citenamefont {Bar},\ and\
  \citenamefont {Schimansky-Geier}}]{Grobmann2014PRL}%
  \BibitemOpen
  \bibfield  {author} {\bibinfo {author} {\bibfnamefont {R.}~\bibnamefont
  {Gro\ss{}mann}}, \bibinfo {author} {\bibfnamefont {P.}~\bibnamefont
  {Romanczuk}}, \bibinfo {author} {\bibfnamefont {M.}~\bibnamefont {Bar}}, \
  and\ \bibinfo {author} {\bibfnamefont {L.}~\bibnamefont {Schimansky-Geier}},\
  }\href@noop {} {\bibfield  {journal} {\bibinfo  {journal} {Phys. Rev. Lett.}\
  }\textbf {\bibinfo {volume} {113}},\ \bibinfo {pages} {258104} (\bibinfo
  {year} {2014})}\BibitemShut {NoStop}%
\bibitem [{\citenamefont {Qiu}\ \emph {et~al.}(2016)\citenamefont {Qiu},
  \citenamefont {Ding}, \citenamefont {Huang}, \citenamefont {Chen},
  \citenamefont {Lu}, \citenamefont {Liu},\ and\ \citenamefont
  {Zhou}}]{Qiu2016PRE}%
  \BibitemOpen
  \bibfield  {author} {\bibinfo {author} {\bibfnamefont {X.}~\bibnamefont
  {Qiu}}, \bibinfo {author} {\bibfnamefont {L.}~\bibnamefont {Ding}}, \bibinfo
  {author} {\bibfnamefont {Y.}~\bibnamefont {Huang}}, \bibinfo {author}
  {\bibfnamefont {M.}~\bibnamefont {Chen}}, \bibinfo {author} {\bibfnamefont
  {Z.}~\bibnamefont {Lu}}, \bibinfo {author} {\bibfnamefont {Y.}~\bibnamefont
  {Liu}}, \ and\ \bibinfo {author} {\bibfnamefont {Q.}~\bibnamefont {Zhou}},\
  }\href@noop {} {\bibfield  {journal} {\bibinfo  {journal} {Phys. Rev. E}\
  }\textbf {\bibinfo {volume} {93}},\ \bibinfo {pages} {062226} (\bibinfo
  {year} {2016})}\BibitemShut {NoStop}%
\bibitem [{\citenamefont {Thampi}\ \emph {et~al.}(2016)\citenamefont {Thampi},
  \citenamefont {Doostmohammadi}, \citenamefont {Shendruk}, \citenamefont
  {Golestanian},\ and\ \citenamefont {Yeomans}}]{Thampi2016science}%
  \BibitemOpen
  \bibfield  {author} {\bibinfo {author} {\bibfnamefont {S.}~\bibnamefont
  {Thampi}}, \bibinfo {author} {\bibfnamefont {A.}~\bibnamefont
  {Doostmohammadi}}, \bibinfo {author} {\bibfnamefont {T.}~\bibnamefont
  {Shendruk}}, \bibinfo {author} {\bibfnamefont {R.}~\bibnamefont
  {Golestanian}}, \ and\ \bibinfo {author} {\bibfnamefont {J.}~\bibnamefont
  {Yeomans}},\ }\href@noop {} {\bibfield  {journal} {\bibinfo  {journal} {Sci.
  Adv.}\ }\textbf {\bibinfo {volume} {2}},\ \bibinfo {pages} {e1501854}
  (\bibinfo {year} {2016})}\BibitemShut {NoStop}%
\bibitem [{\citenamefont {Monin}\ and\ \citenamefont
  {Yaglom}(1971)}]{Monin1971}%
  \BibitemOpen
  \bibfield  {author} {\bibinfo {author} {\bibfnamefont {A.}~\bibnamefont
  {Monin}}\ and\ \bibinfo {author} {\bibfnamefont {A.}~\bibnamefont {Yaglom}},\
  }\href@noop {} {\emph {\bibinfo {title} {{Statistical fluid mechanics vd
  II}}}}\ (\bibinfo  {publisher} {MIT Press Cambridge, Mass},\ \bibinfo {year}
  {1971})\BibitemShut {NoStop}%
\bibitem [{\citenamefont {Wang}(2010)}]{Wang2010JFM}%
  \BibitemOpen
  \bibfield  {author} {\bibinfo {author} {\bibfnamefont {L.}~\bibnamefont
  {Wang}},\ }\href@noop {} {\bibfield  {journal} {\bibinfo  {journal} {J. Fluid
  Mech.}\ }\textbf {\bibinfo {volume} {648}},\ \bibinfo {pages} {183} (\bibinfo
  {year} {2010})}\BibitemShut {NoStop}%
\bibitem [{\citenamefont {Huang}\ \emph {et~al.}(2013)\citenamefont {Huang},
  \citenamefont {Biferale}, \citenamefont {Calzavarini}, \citenamefont {Sun},\
  and\ \citenamefont {Toschi}}]{Huang2013PRE}%
  \BibitemOpen
  \bibfield  {author} {\bibinfo {author} {\bibfnamefont {Y.}~\bibnamefont
  {Huang}}, \bibinfo {author} {\bibfnamefont {L.}~\bibnamefont {Biferale}},
  \bibinfo {author} {\bibfnamefont {E.}~\bibnamefont {Calzavarini}}, \bibinfo
  {author} {\bibfnamefont {C.}~\bibnamefont {Sun}}, \ and\ \bibinfo {author}
  {\bibfnamefont {F.}~\bibnamefont {Toschi}},\ }\href@noop {} {\bibfield
  {journal} {\bibinfo  {journal} {Phys. Rev. E}\ }\textbf {\bibinfo {volume}
  {87}},\ \bibinfo {pages} {041003(R)} (\bibinfo {year} {2013})}\BibitemShut
  {NoStop}%
\bibitem [{\citenamefont {Wang}(2012)}]{Wang2012PoF}%
  \BibitemOpen
  \bibfield  {author} {\bibinfo {author} {\bibfnamefont {L.}~\bibnamefont
  {Wang}},\ }\href@noop {} {\bibfield  {journal} {\bibinfo  {journal} {Phys.
  Fluids}\ }\textbf {\bibinfo {volume} {24}},\ \bibinfo {pages} {045101}
  (\bibinfo {year} {2012})}\BibitemShut {NoStop}%
\bibitem [{\citenamefont {Chakraborty}\ \emph {et~al.}(2014)\citenamefont
  {Chakraborty}, \citenamefont {Wang},\ and\ \citenamefont
  {Klein}}]{Chakraborty2014PRE}%
  \BibitemOpen
  \bibfield  {author} {\bibinfo {author} {\bibfnamefont {N.}~\bibnamefont
  {Chakraborty}}, \bibinfo {author} {\bibfnamefont {L.}~\bibnamefont {Wang}}, \
  and\ \bibinfo {author} {\bibfnamefont {M.}~\bibnamefont {Klein}},\
  }\href@noop {} {\bibfield  {journal} {\bibinfo  {journal} {Phys. Rev. E}\
  }\textbf {\bibinfo {volume} {89}},\ \bibinfo {pages} {033015} (\bibinfo
  {year} {2014})}\BibitemShut {NoStop}%
\bibitem [{\citenamefont {Wang}\ \emph {et~al.}(2013)\citenamefont {Wang},
  \citenamefont {Chakraborty},\ and\ \citenamefont {Zhang}}]{Wang2013PCI}%
  \BibitemOpen
  \bibfield  {author} {\bibinfo {author} {\bibfnamefont {L.}~\bibnamefont
  {Wang}}, \bibinfo {author} {\bibfnamefont {N.}~\bibnamefont {Chakraborty}}, \
  and\ \bibinfo {author} {\bibfnamefont {J.}~\bibnamefont {Zhang}},\
  }\href@noop {} {\bibfield  {journal} {\bibinfo  {journal} {Proc. Combust.
  Inst.}\ }\textbf {\bibinfo {volume} {34}},\ \bibinfo {pages} {1401} (\bibinfo
  {year} {2013})}\BibitemShut {NoStop}%
\bibitem [{\citenamefont {Wang}\ and\ \citenamefont
  {Peters}(2006)}]{Wang2006JFM}%
  \BibitemOpen
  \bibfield  {author} {\bibinfo {author} {\bibfnamefont {L.}~\bibnamefont
  {Wang}}\ and\ \bibinfo {author} {\bibfnamefont {N.}~\bibnamefont {Peters}},\
  }\href@noop {} {\bibfield  {journal} {\bibinfo  {journal} {J. Fluid Mech.}\
  }\textbf {\bibinfo {volume} {554}},\ \bibinfo {pages} {457} (\bibinfo {year}
  {2006})}\BibitemShut {NoStop}%
\bibitem [{\citenamefont {Boffetta}\ and\ \citenamefont
  {Ecke}(2012)}]{Boffetta2012ARFM}%
  \BibitemOpen
  \bibfield  {author} {\bibinfo {author} {\bibfnamefont {G.}~\bibnamefont
  {Boffetta}}\ and\ \bibinfo {author} {\bibfnamefont {R.}~\bibnamefont
  {Ecke}},\ }\href@noop {} {\bibfield  {journal} {\bibinfo  {journal} {Annu.
  Rev. Fluid Mech}\ }\textbf {\bibinfo {volume} {44}},\ \bibinfo {pages} {427}
  (\bibinfo {year} {2012})}\BibitemShut {NoStop}%
\bibitem [{\citenamefont {Germano}(1992)}]{Germano1992JFM}%
  \BibitemOpen
  \bibfield  {author} {\bibinfo {author} {\bibfnamefont {M.}~\bibnamefont
  {Germano}},\ }\href@noop {} {\bibfield  {journal} {\bibinfo  {journal} {J.
  Fluid Mech.}\ }\textbf {\bibinfo {volume} {238}},\ \bibinfo {pages} {325}
  (\bibinfo {year} {1992})}\BibitemShut {NoStop}%
\bibitem [{\citenamefont {Ni}\ \emph {et~al.}(2014)\citenamefont {Ni},
  \citenamefont {Voth},\ and\ \citenamefont {Ouellette}}]{Ni2014PoF}%
  \BibitemOpen
  \bibfield  {author} {\bibinfo {author} {\bibfnamefont {R.}~\bibnamefont
  {Ni}}, \bibinfo {author} {\bibfnamefont {G.}~\bibnamefont {Voth}}, \ and\
  \bibinfo {author} {\bibfnamefont {N.}~\bibnamefont {Ouellette}},\ }\href@noop
  {} {\bibfield  {journal} {\bibinfo  {journal} {Phys. Fluids}\ }\textbf
  {\bibinfo {volume} {26}},\ \bibinfo {pages} {105107} (\bibinfo {year}
  {2014})}\BibitemShut {NoStop}%
\bibitem [{\citenamefont {Chen}\ \emph {et~al.}(2003)\citenamefont {Chen},
  \citenamefont {Ecke}, \citenamefont {Eyink}, \citenamefont {Wang},\ and\
  \citenamefont {Xiao}}]{Chen2003PRL}%
  \BibitemOpen
  \bibfield  {author} {\bibinfo {author} {\bibfnamefont {S.}~\bibnamefont
  {Chen}}, \bibinfo {author} {\bibfnamefont {R.~E.}~\bibnamefont {Ecke}}, \bibinfo
  {author} {\bibfnamefont {G.~L.}~\bibnamefont {Eyink}}, \bibinfo {author}
  {\bibfnamefont {X.}~\bibnamefont {Wang}}, \ and\ \bibinfo {author}
  {\bibfnamefont {Z.}~\bibnamefont {Xiao}},\ }\href@noop {} {\bibfield
  {journal} {\bibinfo  {journal} {Phys. Rev. Lett.}\ }\textbf {\bibinfo
  {volume} {91}},\ \bibinfo {pages} {214501} (\bibinfo {year}
  {2003})}\BibitemShut {NoStop}%
\bibitem [{\citenamefont {Boffetta}\ and\ \citenamefont
  {Musacchio}(2010)}]{Boffetta2010PRE}%
  \BibitemOpen
  \bibfield  {author} {\bibinfo {author} {\bibfnamefont {G.}~\bibnamefont
  {Boffetta}}\ and\ \bibinfo {author} {\bibfnamefont {S.}~\bibnamefont
  {Musacchio}},\ }\href@noop {} {\bibfield  {journal} {\bibinfo  {journal}
  {Phys. Rev. E}\ }\textbf {\bibinfo {volume} {82}},\ \bibinfo {pages} {016307}
  (\bibinfo {year} {2010})}\BibitemShut {NoStop}%
\bibitem [{\citenamefont {Liao}\ and\ \citenamefont
  {Ouellette}(2014)}]{Liao2014PoF}%
  \BibitemOpen
  \bibfield  {author} {\bibinfo {author} {\bibfnamefont {Y.}~\bibnamefont
  {Liao}}\ and\ \bibinfo {author} {\bibfnamefont {N.}~\bibnamefont
  {Ouellette}},\ }\href@noop {} {\bibfield  {journal} {\bibinfo  {journal}
  {Phys. Fluids}\ }\textbf {\bibinfo {volume} {26}},\ \bibinfo {pages} {045103}
  (\bibinfo {year} {2014})}\BibitemShut {NoStop}%
\bibitem [{\citenamefont {Zhou}\ \emph {et~al.}(2015)\citenamefont {Zhou},
  \citenamefont {Huang}, \citenamefont {Lu}, \citenamefont {Liu},\ and\
  \citenamefont {Ni}}]{Zhou2015JFM}%
  \BibitemOpen
  \bibfield  {author} {\bibinfo {author} {\bibfnamefont {Q.}~\bibnamefont
  {Zhou}}, \bibinfo {author} {\bibfnamefont {Y.}~\bibnamefont {Huang}},
  \bibinfo {author} {\bibfnamefont {Z.}~\bibnamefont {Lu}}, \bibinfo {author}
  {\bibfnamefont {Y.}~\bibnamefont {Liu}}, \ and\ \bibinfo {author}
  {\bibfnamefont {R.}~\bibnamefont {Ni}},\ }\href@noop {} {\bibfield  {journal}
  {\bibinfo  {journal} {J. Fluid Mech.}\ }\textbf {\bibinfo {volume} {786}},\
  \bibinfo {pages} {294} (\bibinfo {year} {2015})}\BibitemShut {NoStop}%
\bibitem [{\citenamefont {Kolmogorov}(1962)}]{Kolmogorov1962}%
  \BibitemOpen
  \bibfield  {author} {\bibinfo {author} {\bibfnamefont {A.}~\bibnamefont
  {Kolmogorov}},\ }\href@noop {} {\bibfield  {journal} {\bibinfo  {journal} {J.
  Fluid Mech.}\ }\textbf {\bibinfo {volume} {13}},\ \bibinfo {pages} {82}
  (\bibinfo {year} {1962})}\BibitemShut {NoStop}%
\bibitem [{Note1()}]{Note1}%
  \BibitemOpen
  \bibinfo {note} {See
  {http://damtp.cam.ac.uk/user/gold/datarequests.html}}\BibitemShut {NoStop}%
\bibitem [{Note2()}]{Note2}%
  \BibitemOpen
  \bibinfo {note} {See {https://github.com/lanlankai}}\BibitemShut {NoStop}%
\end{thebibliography}

%

\end{document}